\documentclass[twocolumn]{article}
\usepackage[utf8]{inputenc}

\usepackage{hyperref}

\newcommand{\eq}{\begin{equation}}
\newcommand{\eqx}{\end{equation}}
\newcommand{\eqn}{\begin{eqnarray}}
\newcommand{\eqnx}{\end{eqnarray}}
\newcommand{\f}[2]{\frac{#1}{#2}}

\newcommand{\cor}[1]{\left\langle{#1}\right\rangle}

\newcommand{\sg}{\sigma}

\newcommand{\eps}{\varepsilon}

\newcommand{\qqqq}{\quad\quad\quad\quad}

\title{Explaining the Human Visual Brain Challenge 2019\\
-- receptive fields and surrogate features}

\author{Romuald A. Janik\thanks{e-mail: {\tt romuald.janik@gmail.com}} \\ \\ 
\small 
Jagiellonian University,
Institute of Physics\\\small
ul. {\L}ojasiewicza 11, 
30-348 Krak{\'o}w, 
Poland}
\date{}

\usepackage{fullpage}
\usepackage{graphicx}

\begin{document}

\maketitle

\begin{abstract}
In this paper I review the submission to the {\it Explaining the Human Visual Brain Challenge 2019} in both the fMRI and MEG tracks. The goal was to construct
neural network features which generate the so-called representational dissimilarity matrix (RDM) which is most similar to the one extracted from fMRI and MEG data upon viewing a set of images. I review exploring the optimal
granularity of the receptive field, a construction of intermediate surrogate features using Multidimensional Scaling and modelling them using neural network
features. I also point out some peculiarities of the RDM construction which
have to be taken into account.
\end{abstract}

\section{Introduction}

The {\it Explaining the Human Visual Brain Challenge 2019} organized within the Algonauts project \cite{ALGONAUTS}, has as its aim increasing our understanding of what visual features are processed and encoded by the brain. The challenge specifically concentrates on fMRI data from the early visual cortex (EVC) and inferior temporal cortex (IT) in the fMRI track of the competition and on MEG data from the early and late stages of visual processing
in the MEG track of the competition. In the following I will use the above
acronyms as well as EARLY and LATE targets for the MEG track.

\begin{figure}[h]
\centering
\includegraphics[height=3cm]{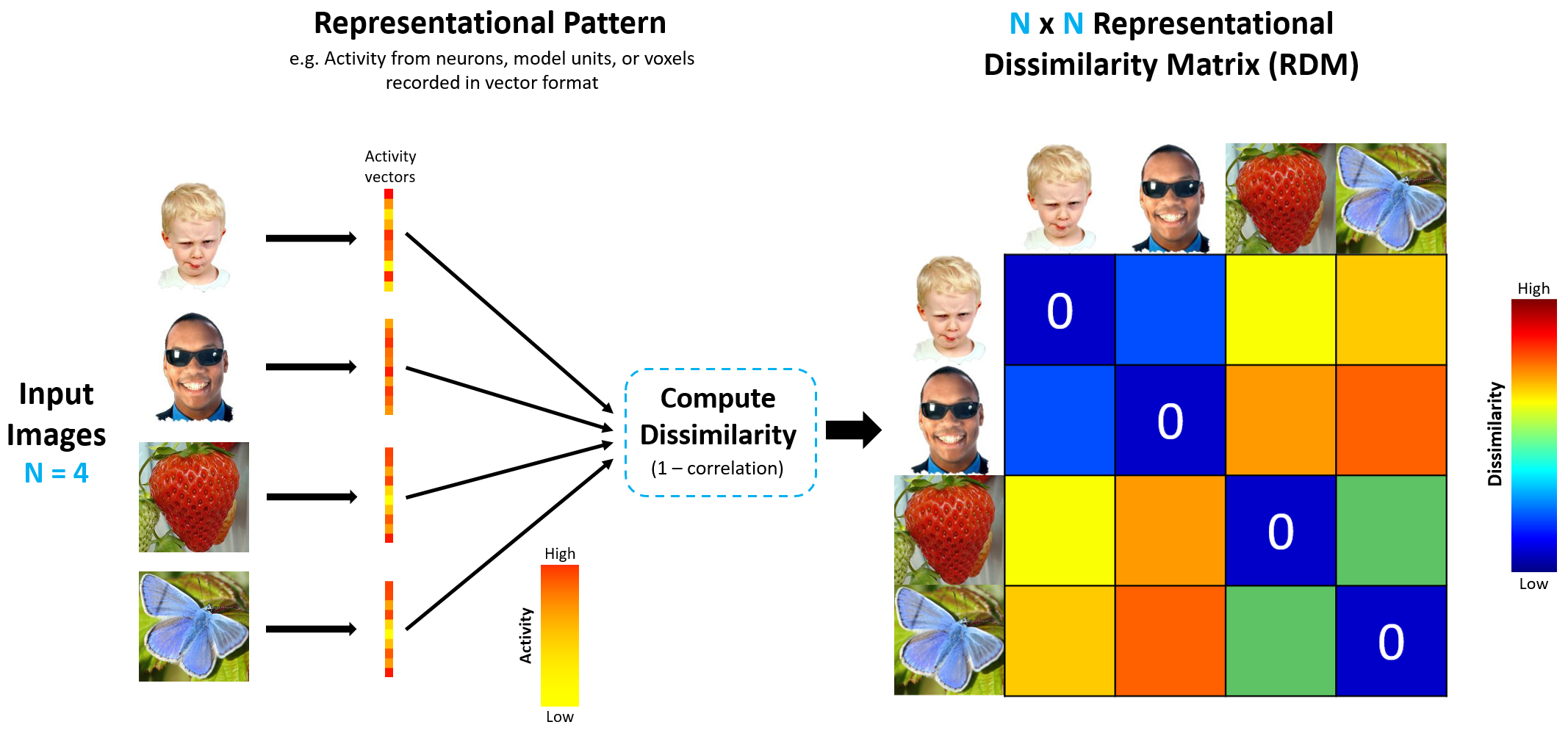}
\caption{The representational dissimilarity matrix (RDM). (figure from \cite{FIGS})}
\label{fig.rdm}
\end{figure}

The competition adopts the Representational Similarity
Analysis framework \cite{RSA} where one first constructs a so-called representational dissimilarity matrix (RDM) out of brain data, whose entries are $1-Pearson\ correlation$ between the neuronal activities/signals corresponding to two images (see Fig.~\ref{fig.rdm}).

\begin{figure}[h!]
\centering
\includegraphics[height=4.5cm]{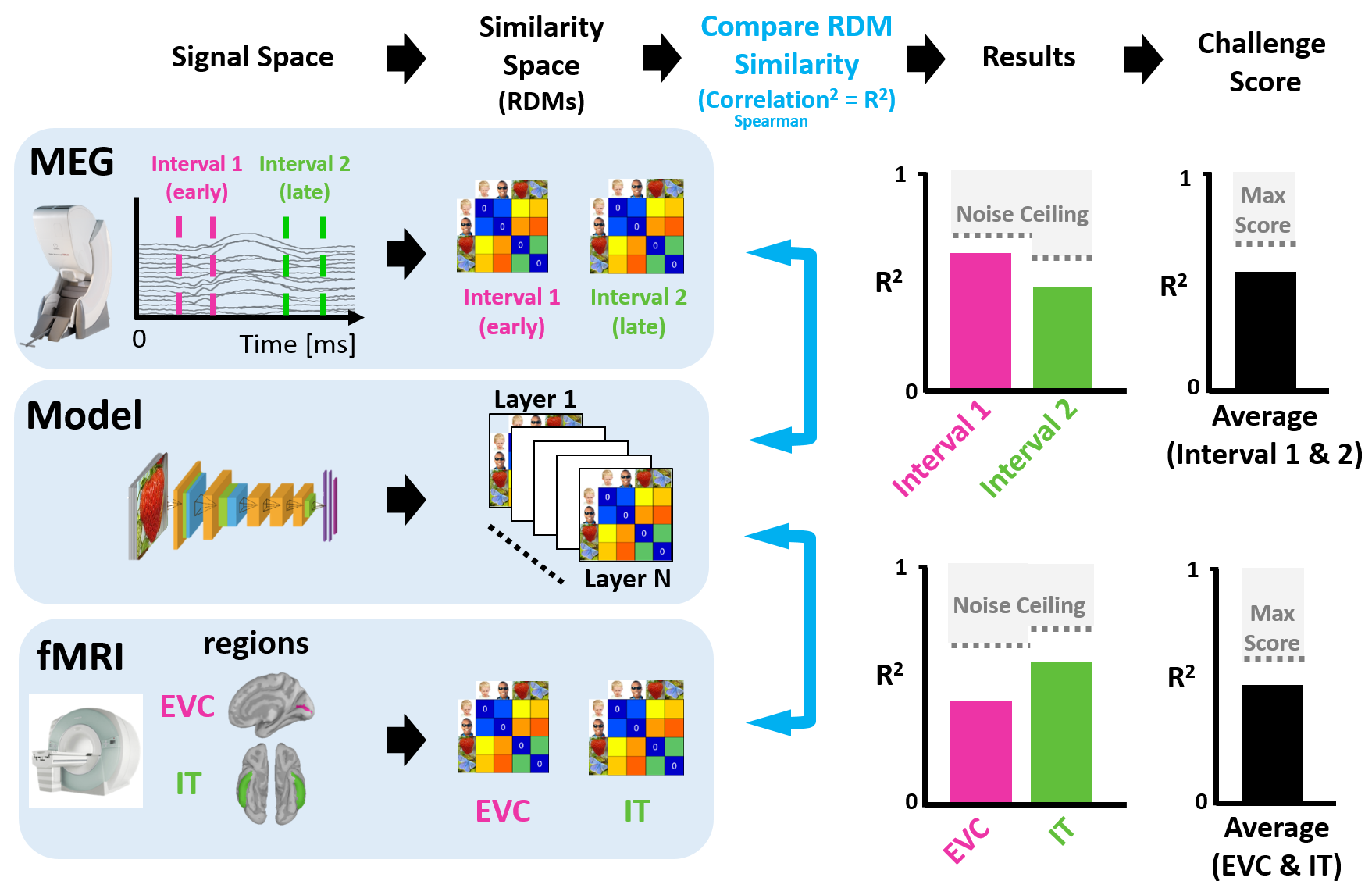}
\caption{The general framework of the competition. (figure from \cite{FIGS})}
\label{fig.competition}
\end{figure}

The goal of the competition is to construct features (e.g. out of Deep Neural Network activations) so that the resulting dissimilarity matrix is closest,
as measured by Spearman's correlation, to the one extracted from the brain
averaged over 15 subjects (see Fig.~\ref{fig.competition}). There are two training datasets, with 118 and 92 images respectively and and a test set with 78 images. For more details see \cite{ALGONAUTS} and \cite{WEBSITE}.

Below I first discuss some very general ingredients of the approach: some
rather unexpected properties of the RDM's, determining the optimal coarse grained receptive field and the methodology for feature selection. Then I outline the approach using surrogate MDS features and ICA components for the IT and LATE targets, and the models for the EVC and EARLY targets.
Since this is a write-up of a competition entry, some clarity of the modelling approach is unfortunately sacrificed for tweaking performance.

\section{Peculiarities of the RDM construction}
\label{s.pathologies}

The Representational Dissimilarity Matrices (RDM) by construction have two rather unexpected and somewhat unwelcome features. Firstly, they can miss a very strong discriminative signal as long as it is correlated. Secondly,
they are influenced by the presence of irrelevant uninformative features.

\begin{figure}
\centering
\includegraphics[width=0.45\textwidth]{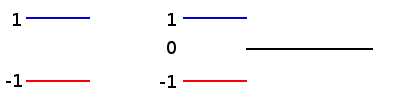}
\caption{Features for two images which are not captured by the RDM (left). Adding a vector of uninformative features makes the
correlated relevant features visible for RDM (right).}
\label{fig.rdmcorr}
\end{figure}

The dissimilarity between features $x$ and $y$ representing two images is
defined by
\eq
1 - R(x,y) = 1- \f{(x-\cor{x})(y-\cor{y})}{\sg_x \sg_y}
\eqx
where $R(x,y)$ is Pearson's correlation coefficient between the two feature vectors. Suppose that the two images are completely dissimilar with the features $x$ being all $+1$, and the features $y$ being all $-1$ (as shown in Fig.~\ref{fig.rdmcorr} (left)).
Since the Pearson correlation coefficient subtracts the mean of the feature vector, the difference
between the two images will be completely lost.
In a more realistic setting when
\eq
x_i = +1 +\eps_i \qqqq y_i = -1 +\tilde{\eps}_i
\eqx
with some noise vectors $\eps_i$ and $\tilde{\eps}_i$, the computed 
dissimilarity will only be influenced by the correlation of the noise signals
and not by the relevant discriminating features.

Suppose that we concatenate to our features a vector of $N$ irrelevant
features, which are always zero irrespective of the image (as shown in Fig.~\ref{fig.rdmcorr} (right)).
Now the relevant features $+1$ and $-1$ will start to contribute to the
dissimilarity. In the limit $N\to \infty$, this corresponds to using
a cosine dissimilarity
\eq
1- \f{x\cdot y}{|x| |y|}
\eqx
for the original features.

For the neural network features, the global signal (e.g. $\cor{x}$ above)
carries important information and should not be subtracted out. Hence we always add a constant vector to our constructed features\footnote{95\% overall length.} for all entries to the competition. This strongly influences the results.
We will refer to this constant as $level$.

It may well be that uninformative voxels may also appear in the brain RDM's,
so the above may not be completely artificial. However the sensitivity
of the dissimilarity to such uninformative voxels and the downplaying of correlated signals is rather unwelcome. Fortunately, the two properties
work in opposite directions but this adds another degree of freedom to the setup.
One way of correcting it would
be treat $\cor{x}$ as an average over all images for each feature individually.
However, the definition of the RDM is of course completely fixed throughout
the competition.

\section{Effective receptive fields}

The spatial size of the outputs of the different convolutional layers
vary significantly (e.g. for {\tt resnet50} we have $56\times 56$, $28\times 28$, $14\times 14$ and $7\times 7$). This corresponds to looking at the images
at different "granularities". Of course, the features (channels) for each layer
also differ in their "complexity" going intuitively from low-level features
to high-level ones.
It is interesting to investigate what is the natural granularity (and whether
something like this exists)
for brain regions/processing stages appearing in the competition,
disentangled from the low-level/high-level feature characteristics.

\begin{figure}
\centering
\includegraphics[width=0.45\textwidth]{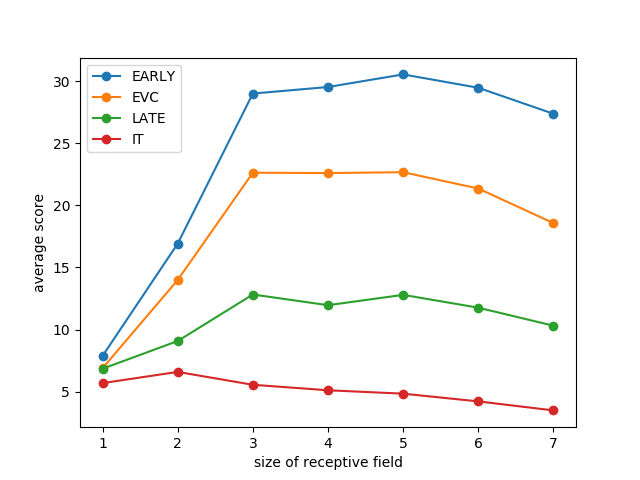}
\caption{Average score on the convolutional layers of {\tt resnet18}, {\tt resnet34} and {\tt resnet50} (with constant $level=0.5$) on the 118 image dataset.}
\label{fig.receptive}
\end{figure}

To this end, we can transform the outputs from the convolutional layers through
a max-pooling operation leading to a specified $k \times k$ output size\footnote{This can be done using the {\tt adaptive\_max\_pool2d} operation in PyTorch.}.
The results for the average score on the 118 image dataset of the transformed {\tt resnet} family convolutional layers for different modalities is shown in Fig.~\ref{fig.receptive}.  
Hence we pick out a $5\times 5$ grid for the EVC, EARLY and LATE targets, while IT favours a very coarse $2\times 2$ grid. Some random trials on
the test set also support this conclusion.

It is interesting to note that if one would use average pooling with the
same settings, the results would be much worse -- single digit scores with the same settings as in Fig.~\ref{fig.receptive}. Thus the brain seems to pick
out just the strongest signals in appropriate parts of the image.

In the following we always preprocessed the convolutional layers in this
way, normalizing the resulting outputs so that the mean of nonzero elements
would be equal to unity.

\section{Methods for feature selection}
\label{s.fsel}

In the competition entries we frequently screened features for their relevance
and eliminated the irrelevant ones. Due to the fact that the dissimilarity
matrix is a very global way of summarizing features, it seems rather difficult
to define a relevant feature without a particular context of other features.
Hence, we always define relevance in the context of some prescribed
set of features. We evaluate the score on this reference set, and
compute the difference w.r.t this score when we erase a feature from the set\footnote{By replacing it with the constant $level$ of section~\ref{s.pathologies}.}, or when we add an external feature to this set\footnote{In the context of convolutional features, we always erase or add
an output channel as a whole.}.

The rather complicated prescriptions outlined below are motivated by trying
to isolate genuine relevant or irrelevant features, so that they would have a good chance of generalizing to new data. However this is by no means guaranteed to work, and only works to some extent.

We used two possible ways of scoring features:
\begin{enumerate}
\item[A] For each of the 15 subjects we \emph{individually} evaluate
the reference score and the modified score (with an added or erased feature). We then take the mean, as well
as the $z$-score of the 15 differences.
\item[B] We randomly choose 30 subsets of around $1/4$ images and use these
for the reference and modified scores. Similarly we take the mean and $z$-score of the 30 differences\footnote{For some inessential historical reason, we did that
within an overall 5-fold cross-validation.}.
\end{enumerate}

For features constructed from some fitting process, as in section~\ref{s.mds}, we use option A on the CV test folds, as well as on predictions on the other dataset (e.g. 118 if the model was trained on 92 images and vice-versa). We will refer
to this procedure later as {\bf MDS pruning}.

For adding features we used a slightly more complicated process. First we used
10 different splits of the dataset into 5 parts. For each part we computed the
gain in the score resulting from adding a feature. From these 50 differences
we computed the mean. As a set of candidate features to add we picked those which had positive mean for both the 118 and 92 image dataset. Then we added successively the best features and kept those which still had a positive mean.
We will refer to this procedure later as {\bf Feature Adding}.

\begin{figure}[t]
\includegraphics[width=0.4\textwidth]{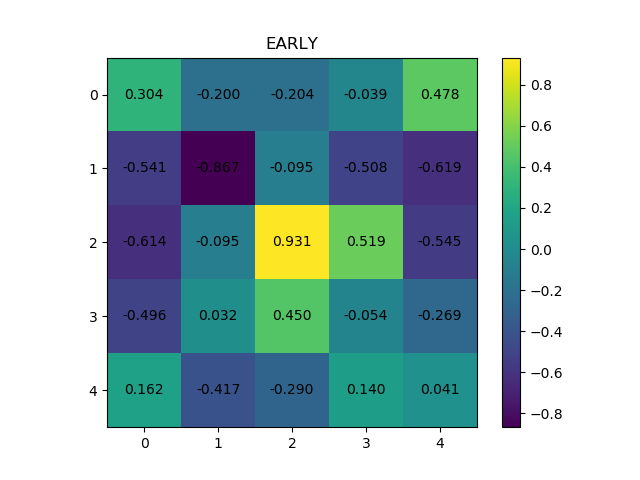}\hfill%
\includegraphics[width=0.4\textwidth]{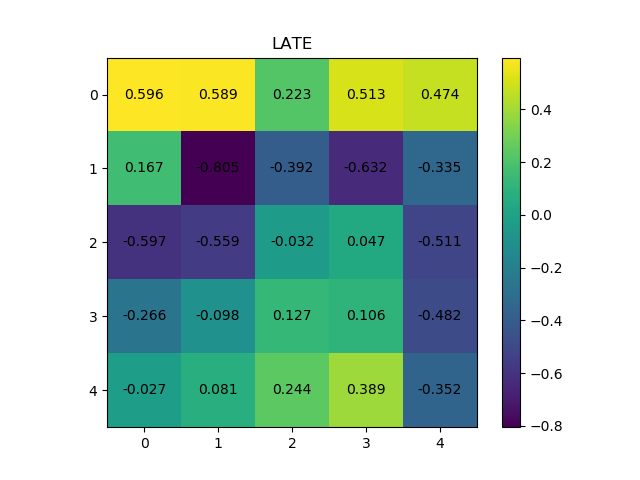}
\caption{$z$-scores when dropping part of the receptive field on the 118 image dataset modelled by {\tt maxpool2} of {\tt vgg19}. Positive numbers indicate that scores increase when that part is erased.}
\label{fig.spatial}
\end{figure}

An interesting outcome of quantifying feature importance is the relevance of the
$5\times 5$ spatial grid when modelling EARLY, LATE (and to some extent EVC).
Applying algorithm A by dropping all channels in a particular grid position,
we get a $5\times 5$ matrix of $z$-scores (shown in Fig.\ref{fig.spatial} for the LATE and EARLY processing stages on the 118 image dataset modelled by the {\tt maxpool2} layer of {\tt vgg19}). We henceforth drop the corners of the EARLY stage features. 
We would also drop the top and bottom row of the LATE stage features, but
in the end we adopted a different approach in this case.
Both conclusions were confirmed on the test set.
We also found that dropping corners on the EVC also gave a perfomance boost.

\section{Surrogate MDS features and the IT and LATE features}
\label{s.mds}

In order to model the dissimilarity matrices for the IT and LATE part of the challenge, we decided to construct first completely abstractly features which
could reproduce well the dissimilarity matrix. These features do not have any
\emph{a-priori} interpretation and are defined only for a given training dataset,
hence we call them surrogate features.
Then we fit various models to reproduce these abstract features from 
the neural network features. These models can then be applied to the test dataset to yield the predicted features.

The idea of the construction is to use Multidimensional Scaling (MDS) embedding constructed out of the average (across subjects) dissimilarity matrix\footnote{More precisely we find that the $4^{th}$ power of the mean RDM works best with the MDS algorithm.}. 
MDS aims to find a low dimensional embedding which preserves as
well as possible the distances (related here to dissimilarities) between
data points (images). Of course, it does not follow that the resulting 
coordinates will lead to the original dissimilarity matrix, but experiments on the 118 dataset gave sufficiently high scores so that
trying to develop a custom embedding algorithm was not worthwhile
during the competition.

Since the MDS algorithm is probabilistic, we fix 10 random seeds and construct 10-dimensional MDS embeddings giving in total 100 features.
For the 118 image dataset, these features allow theoretically to obtain a score of around 77\%. In practice the score will be lower as we have to express (approximately) these
features in terms of DNN features. Due to the small number of images and huge number of DNN features we have to impose strong regularization or sparsity constraints. Also, in order to reduce somewhat the number of DNN features, we fit the model to each layer in turn and then combine the results using ridge regression.

For the IT target, we use ridge\footnote{{\tt RidgeCV} from sklearn.} regression, as well
as orthogonal matching pursuit OMP(6) (with 6 nonzero components per layer) trained on the 118 dataset and OMP(7) trained on the 92 image dataset.
All these models are trained on layers of the {\tt resnet50} network (with convolutional layers reduced to $2\times 2$).
We concatenate the features giving a vector of length 300 for each image.
We then use the {\bf MDS pruning} procedure by imposing positivity on the 118 dataset. The final answer is extended with a constant level of $1$ of section \ref{s.pathologies} (the precise value is not important as long as it is away from zero). This yields a score of 19.42 on the test set.

From some earlier attempts we found that ICA features work rather well on the IT (but worse than above). Hence, we concatenate the previous 300 features with 75+75 ICA components of {\tt block1} and {\tt block3} respectively\footnote{Since ICA works worse for the IT target, we shrink them by multiplying by 0.25 and 0.5 respectively.}, from {\tt resnet34}. ICA is fitted on the 118 and 92 datasets concatenated together. This raises the score to {\bf 20.77} for the IT target on the test set.

For the LATE target, a cross validation study of ridge and OMP does not give good results. However, rather surprisingly a nonlinear model -- gradient boosting regression
works rather well. Since we expect the data to be very noisy and prone to overfitting, we use only very small trees (with $n_{estimators}=5$) and 
Huber loss. We concatenate GBR(5) fitted on the 118 dataset MDS and GBR(5) fitted
on the 92 dataset MDS (again on the {\tt resnet50} network with convolutional layers reduced now to $5\times 5$)).
We then use the {\bf MDS pruning} procedure by eliminating features which are
negative on both datasets. Again we extend the set with a constant level of $1$.
This yields a score of 57.10. Eliminating a slightly smaller number of features by raising the elimination threshold to 0.05 gives the final score {\bf 57.38} for the LATE target on the test set.

\section{The EARLY and EVC features}

The models for EARLY and EVC are quite unsophisticated and involve a choice of DNN layer, reduction to $5 \times 5$, eliminating corners of the receptive field, eliminating some features using some pruning procedure and then enhancing the most relevant ones.
One can make a very simple model for the LATE target along these lines which gives 33.24 on the test set (just after pruning). However,
for our final model we adopted the approach using MDS features.
The details for the EARLY and EVC targets are given below.

For the EARLY target, we use the {\tt maxpool2} layer of {\tt vgg19}. For the pruning step we use algorithm A of section~\ref{s.fsel} eliminating features with negative $z$-scores bigger than 0.15 on either dataset. We than used the {\bf Feature Adding} procedure on the same layer, and concatenated these features with a multiplier $4$. The constant level was here set to 0.5. This yields a score of {\bf 46.91} for the EARLY target on the test set.

For the EVC target, we use the {\tt block2} layer of {\tt resnet18}. For the pruning step
we use algorithm B of section~\ref{s.fsel} eliminating 1/4 of the worst features (according to their $z$-scores). We than used the {\bf Feature Adding} procedure on the same layer, and concatenated these features with a multiplier $2$. The constant level was here set to 0. This yields a score of 28.29.  Performing {\bf Feature Adding} with the {\tt maxpool2} layer of {\tt vgg19} and concatenating with multiplier 0.5 gives a slight boost to {\bf 28.40} for the EVC target on the test set.

\section{Conclusions}

The key difficulty in the challenge is the relatively small number of images in comparison with the enormous number of potential DNN features. In addition,
the three datasets are quite distinct, far from satisfying an i.i.d. assumption, so that there is a huge potential for overfitting. E.g. direct optimization of the score as a function of feature weights using some gradient free optimizer did not generalize well when
doing cross-validation.
The same happened for some approaches which had the potential of working for much larger datasets like a Siamese network on top of the DNN features. In view of the above we concentrated on rather simple models. 

In addition, unfortunately cross-validation was not always a reliable
indicator of performance on the test set. An extreme example
was a submission which (erronously) added \emph{worst} features from 
the remaining layers of {\tt resnet18} in the {\bf Feature Adding} step to EVC, which gave {\bf 32.68} on the test set.

Some properties seemed quite robust, however, like the optimal effective granularity of the receptive field is not very large (around $5 \times 5$ or $2 \times 2$), as well as the fact that roughly second level layers (like {\tt block2} or {\tt maxpool2}) seemed to be the best starting point for the early stages EVC and EARLY targets.

The insensitivity of the RDM construction to correlated features as well as the fact that it is influenced by completely uninformative features suggests that it would be better to modify its definition e.g. by subtracting out the featurewise average over the dataset.

The approach which worked best for the late stages (IT and LATE targets), used
surrogate features constructed out of MDS embedding of the (transformed) dissimilarity matrix and then various models for fitting these features.
If one would want to seriously proceed along this route, then one
should develop a different embedding algorithm which would be 
directly optimized for reproducing the given RDM.
Of course, given the original brain fMRI and MEG data, one could directly
try to fit these data (after some appropriate preprocessing) with the neural network features. This would have the additional benefit of
explaining some concrete real brain feature.
We intend to investigate this approach in the future.

\smallskip

\noindent{}{\bf Acknowledgments.}
This work was done in preparation for the Foundation for Polish Science (FNP) project \emph{Biologically inspired Artificial Neural Networks} POIR.04.04.00-00-14DE/18-00. I would like to thank the organizers 
for a very stimulating challenge.


\begin{thebibliography}{XX}
\bibitem{ALGONAUTS} R. M. Cichy, G. Roig, A. Andonian, K. Dwivedi, B. Lahner, A. Lascelles, Y. Mohsenzadeh, K. Ramakrishnan, A. Oliva. {\it The Algonauts Project: A Platform for Communication between the Sciences of Biological and Artificial Intelligence.} arXiv: \href{https://arxiv.org/abs/1905.05675}{1905.05675}
\bibitem{RSA} N. Kriegeskorte, M. Mur, P.A. Bandettini, {\it Representational similarity analysis-connecting the branches of systems neuroscience.} Frontiers in systems neuroscience, 2, 4 (2008).
\bibitem{FIGS} \url{http://algonauts.csail.mit.edu/rsa.html}
\bibitem{WEBSITE} \url{http://algonauts.csail.mit.edu/challenge.html}
\end{thebibliography}
\end{document}